\documentclass[useAMS,usenatbib,usegraphicx
]{mn2e}
\usepackage{amssymb,deluxetable}
\usepackage{aas_macros}
\newcommand{\header}[1]{\multicolumn{1}{c}{\textrm{#1}}}
\renewcommand{\i}{\,{\sc i}}
\newcommand{\ii}{\,{\sc ii}}

\newcommand{\iv}{\,{\sc iv}}
\newcommand{\Teff}{$T_{eff}$}
\newcommand{\lgg}{\rm{log}\,$g$}

\newcommand{\kms}{km\,s$^{-1}$}
\newcommand{\pz}{PZ~Mon}
\newcommand{\vr}{$v_{\rm rad}$}

\newcommand{\vm}{$v_{\rm macro}$}
\newcommand{\vs}{$v\sin{i}$}
\newcommand{\Ha}{H$\alpha$}

\title{Evolutionary status of the active star \pz}

\author[Yu.V.Pakhomov,N.N.Chugai,N.I.Bondar,N.A.Gorynya,E.A.Semenko]{
Yu.~V.~Pakhomov,$^1$
N.~N.~Chugai,$^1$
N.~I.~Bondar,$^2$
N.~A.~Gorynya,$^{1,3}$
E.~A.~Semenko,$^4$
 \\
$^1$Institute of Astronomy, Russian Academy of Sciences, Pyatnitskaya 48,
119017, Moscow, Russia\\
$^2$Crimean Astrophysical Observatory, Nauchny, Crimea, 2984009, Russia\\
$^3$Lomonosov Moscow State University, Sternberg Astronomical Institute,
Universitetskij prospekt, 13, Moscow 119991, Russia\\
$^4$Special Astrophysical Observatory of Russian Academy of Sciences, Russia
}

\begin{document}
\maketitle

\begin{abstract}
We use original spectra and available photometric data to recover parameters of
the stellar atmosphere of \pz, formerly referred as an active red dwarf. 
The derived effective temperature \Teff=4700~K and gravity \lgg=2.8 suggest that
\pz\ is a K2III giant. Stellar atmosphere parameters (\Teff\ and \lgg) alongside
with the evolutionary tracks are used to estimate the stellar mass of
$\approx1.5~M_\odot$ and the radius of $\approx7.7~R_\odot$. The angular radius
derived by the infrared flux method when combined with the linear radius
suggests the distance of $250\pm70$~pc, a factor 2.5 smaller than that suggested
by the {\it Hipparcos} parallax. The red giant status of \pz\ is confirmed by
the carbon and nitrogen abundance. The spectrum reveals pronounced He\i\ 5876
\AA\ absorption and \Ha\ emission indicating the robust chromosphere.
The {\it IUE} spectrum is found to contain transition layer emission line of 
C\iv\ 1550 \AA. The C\iv\ and X-ray luminosities turn out typical of RS~CVn
stars. The extended set of available photometric data confirms the period of
34.14~days presumably related to the stellar rotation. We found variations of
the radial velocity with the amplitude of $\approx 8$~\kms\ which could be 
caused by the orbital motion.

\end{abstract}


\section{Introduction}

\pz\ (HD\,289114, $V\approx9$~mag) is K2Ve star rapid irregular variable of
F-M/Fe-Me type \citep{2009yCat....102025S} (see also SIMBAD). Now this
classification seems an obsolete which reflects the history of \pz\ study rather
then the present day situation. First, however, we have to recap major
observational facts. Pronounced variable Ca\ii\ and hydrogen emissions were
found in several spectra taken between 1948 and 1954 which has prompted to
classify \pz\ as a flare K2-dwarf at the distance of $\sim 30$~pc
\citep{1955BOTT....2m..36M}. The light curve of \pz\ recovered from Harward
archive between 1899 and 1954 demonstrated flare-type variations
\citep{1955BOTT....2m..39G} although it was stressed that the star had the
relatively high visible magnitude for variables of this class. Later on \pz\
was referred to as UV~Cet type star \citep{1958MmSAIS..2...29P,
1999A&AS..139..555G}, although attribution to RW~Aur type was suggested as well
\citep{1959SvA.....3..832A}. The first distance estimate (30~pc) has been
reduced to 16-21~pc  \citep{1982A&AS...47..471G, 1989MNRAS.238..709S} with the
lower limit preferred by \cite{1999A&AS..139..555G}.
 
The compiled light curve of \pz\ on the interval of 90~yr recovered from archive
plates \citep{1995A&AS..111..259B} shows long-term variations with the total $B$
amplitude $\sim0.8$~mag and the possible period of $\sim50$~yr. These variations
have been interpreted as a result of the activity of a spotted red dwarf 
\citep{1996BCrAO..93...95B, 2006A&AT...25..247A}. \pz\ shows also short-term
flux variations with the total amplitude $\approx0.04$~mag and period of about
34~d \citep{2007OAP....20...14B}; they were attributed to the rotational flux
modulation of the spotty star with the rotation period of 34~d.

The new history starts with the {\em Hipparcos} parallax data which suggested
the distance of 1.4~kpc and the absolute magnitude $M_V\approx 0.2$~mag
\citep{1998IBVS.4580....1S}. This combined with the known X-ray flux
\citep{1986MNRAS.219..225A} prompted to make an assumption that \pz\ is an
active giant, probably of RS~CVn type \citep{1998IBVS.4580....1S}. To support
the giant status \citet{1998IBVS.4580....1S} compared the spectrum of \pz\ with
the red dwarf (HD~32147, K4V) and RS~CVn type giant $\lambda$~And (G8III-IV) and
concluded that the giant matched better. However, later on it became clear 
that metallicity of both comparison stars was essentially non-solar:
HD~32147 has enhanced metallicity \citep[ \mbox{[Fe/H]$\approx$}0.2...0.3,
][]{2011A&A...531A.165P, 2012A&A...541A..40M} while $\lambda$~And is metal poor
\citep[\mbox{[Fe/H]=}-0.56, ][]{2008AJ....135..209M}. These facts weaken
spectral arguments in favour of giant status of \pz. 

The change in the \pz\ status raises several urgent questions: 1) Whether \pz\
is a giant from the spectroscopic point of view? 2) If does, what are the
stellar parameters and distance to \pz? 3) Whether the chromosphere and corona
are typical to active red giants of RS~CVn type? 4) Are there any evidence
that \pz\ is a binary? The point is that even the revised {\em Hipparcos}
parallax ($\pi$=1.57$\pm$1.01~mas) \citep{2007A&A...474..653V},  with the
distance of 640$^{+1150}_{-250}$~pc -- lower than adopted by
\citet{1998IBVS.4580....1S} -- admits a broad range of the luminosities. The
answer to the posed questions constitute the subject of the present paper. Our
motivation is also strengthen by the fact that \pz\ has the longest (over
100~yr) record of the photometric variability with the large amplitude
($\sim0.8$~mag) of the long-term variation among RS~CVn stars. 

Below we address the \pz\ status issue concentrating on the detailed analysis of
the spectral and photometric data. Bearing in mind that \pz\ might be a binary 
\citep{1998IBVS.4580....1S} we explore also the radial velocity variations. In
Sec.~\ref{sec-phot} we analyse photometric data to determine the effective 
temperature and the angular radius and study the photometric variability using
more extended photometric set than before. In Sec.~\ref{sec-spec} the spectrum
of \pz\ will be analysed and parameters of stellar atmosphere and elemental
abundances will be recovered. We then use the effective temperature and gravity 
alongside with evolutionary tracks to obtain the stellar parameters,
specifically, mass, radius, luminosity, and distance (Sec.~\ref{sec-param}). 
In Sec.~\ref{sec-chrom} we consider signatures of the chromosphere in the
optical and {\it IUE} spectra and then finally discuss the nature of \pz.

\section{Photometry}
\label{sec-phot}

\subsection{Effective temperature}

\begin{figure}
\centering
\resizebox{0.85\hsize}{!}{\includegraphics[clip]{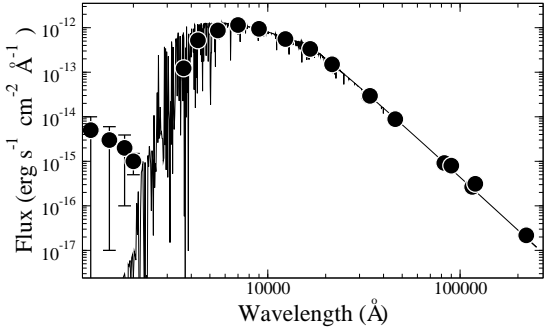}}
\caption{Spectral energy distribution of PZ Mon ({\it dots}) with the
overplotted best fit model ({\it line}).}
\label{fig:sed}
\end{figure}

Photometric data in $UBVRI$ Jonhson-Cousins system \citep{2006A&AT...25..247A}
show variability of \pz\ during 1992-2004 with the amplitude of 0.2-0.3~mag on
the time scale of the hundreds days. The average color indices are
$U-B$=1.05$\pm$0.04~mag, $B-V$=1.17$\pm$0.02~mag, $V-R$=1.18$\pm$0.02~mag,
$V-I$=1.73$\pm$0.03~mag. The IR color indices are $V-J$=2.25$\pm$0.03~mag,
$V-H$=2.83$\pm$0.03~mag, $V-K$=2.99$\pm$0.03~mag, and $V-L$=3.03$\pm$0.04~mag
\citep{1978A&AS...34..477M, 2003yCat.2246....0C}. We used a grid of synthetic
colors \citep{2003IAUS..210P.A20C} to find the best fit of all the available
colors with the interstellar reddening taken into account according to  
\citet{1990ARA&A..28...37M}. The best fit at the 3$\sigma$ level is found for
two temperatures \Teff=4500~K with $E(B-V)$=0.09~mag, and \Teff=4750~K with
$E(B-V)$=0.19~mag.

\begin{table}
\label{tab:irfm}
\caption{Effective temperature and angular diameter of \pz\ derived from the
photometric data by infrared flux method.}
\centering
\begin{tabular}{rrr}
\hline
  \header{E(B-V)} &
  \header{\Teff, K} &
  \header{$\theta$, mas} \\
\hline
0.00 & 4640$\pm$80 & 0.28$\pm$0.01 \\
0.03 & 4670$\pm$80 & 0.28$\pm$0.01 \\
0.06 & 4700$\pm$80 & 0.28$\pm$0.01 \\
0.10 & 4750$\pm$80 & 0.29$\pm$0.01 \\
\hline
\end{tabular}
\end{table}

The infrared flux method (IRFM) \citep{1977MNRAS.180..177B} permits us to
determine the angular diameter and the effective temperature in a more reliable
way. The empirical spectral energy distribution of \pz\ is recovered combining
UV flux from IUE data \citep{2000Ap&SS.273..155W}, optical flux from the $UBVRI$
photometry, and near IR $2MASS$ data \citep{2003yCat.2246....0C}. We use also
far IR flux of 0.2159$\pm$0.00892~Jy at 9\,$\mu$  from the ARARI survey
\citep{2010A&A...514A...1I}, 8.28\,$\mu$ flux of 0.2093~Jy
\citep{2003yCat.5114....0E}, 12\,$\mu$ flux of 0.15~Jy
\citep{1999A&AS..139..555G}, and IR fluxes from WISE mission
\citep{2010AJ....140.1868W}: $W1({3.35\,\mu})$=6.177$\pm$0.088 mag,
$W2({4.6\,\mu})$=6.078$\pm$0.036 mag, $W3({11.6\,\mu})$=6.073$\pm$0.015 mag,
and $W4({22.1\,\mu})$=5.927$\pm$0.046 mag. The IR fluxes are only weekly
affected by the reddening and temperature, which make them good indicators  of
the angular diameter $\theta = 2(f_\lambda^{obs}/F_\lambda)^{1/2}$, where 
$f_\lambda^{obs}$ is the observed flux and $F_\lambda$ is the flux at the
stellar photosphere. We proceed in a standard way using the iteration procedure
and first guess for \Teff. The theoretical spectrum
\citep{2003IAUS..210P.A20C} for the adopted first guess $T_{eff}=4700$~K
is used to derive $\theta$. This value and the integrated observed flux $F$ is
used to obtain new \Teff\ value from the relation
$T_{eff}=(4F/\theta^2\sigma)^{1/4}$, where $\sigma$ is the Stefan-Boltzmann
constant. The iterations are repeated until the convergence. The spectral
energy distribution for Kurucz model with $T_{eff}=4700$~K and $E(B-V)=0.06$~mag
is shown in Fig.~\ref{fig:sed}. The UV flux with high brightness temperature is
strongly enhanced compared to the photospheric flux and undoubtedly is due to
chromosphere of \pz. The results of the \Teff\ determination by IRFM are
shown in Table~\ref{tab:irfm}. Starting from the first column the table gives
adopted reddening, effective temperature, and angular diameter. We made an
attempt to derive the reddening in the framework of IRFM but found that this
procedure was not unequivocal. We therefore show the temperature estimates for
the several reddening values. The preferred value $E(B-V)=0.06$~mag is
determined by the equality between \Teff\ inferred by IRFM and spectroscopic
method. Remarkably, the angular diameter is insensitive to the choice of the
reddening.

\subsection{Photometric variability}

\begin{figure}
\centering
\resizebox{0.95\hsize}{!}{\includegraphics[clip]{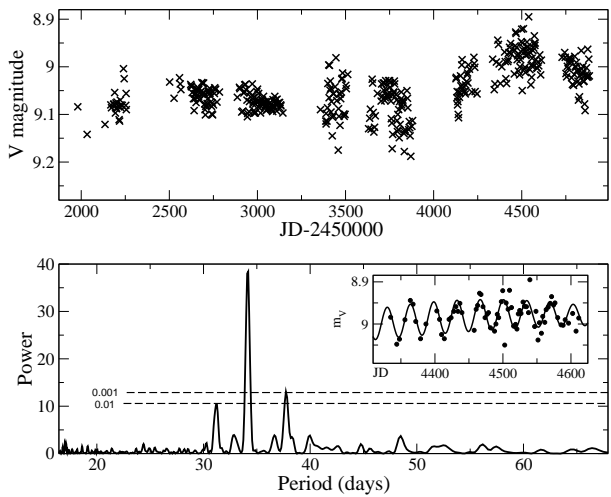}}
\caption{Light curve of \pz\ according to ASAS-3 data ({\it upper} panel) and 
power spectrum of the Lomb periodogram ({\it lower} panel). Horizontal lines
show the significance levels. The {\it inset} shows a fragment of the ASAS-3
data record with overplotted 34.14~d oscillations.}
\label{fig:asas}
\end{figure}

The extended set of ASAS-3 data
\citep[][http://www.astrouw.edu.pl/asas/]{1997AcA....47..467P} between 2001 and
2009 is used to study the \pz\ variability (Fig.~\ref{fig:asas}). The reported
data uncertainty is about 0.04~mag, although the actual error seems to be
somewhat smaller. Indeed, the flux variation during JD 2454335--2454611 (Aug
2007 - May 2008) shows the apparent periodic signal with the amplitude of 
$\approx0.05$~mag and period of about 34~d (Fig.~\ref{fig:asas}, inset).
We analyzed all the available data sets by the Lomb method to recover 
significant peak in the periodogram (Fig.~\ref{fig:asas}) at 34.14 days. Two
smaller side peaks have caused by the gaps in the data set. The found period
well coincides with the period value obtained earlier
\citep{2007OAP....20...14B} and identified with the rotational period.

\section{Spectroscopy}
\label{sec-spec}

\subsection{Observations and spectrum overview}
\label{spectrum}

Two spectra of \pz\ were obtained 2012 November 4/5 with NES2 echelle
spectrograph installed in Nasmyth focus at 6-m telescope of Special
Astrophysical Observatory of Russian Academy of Sciences. Two one hour exposures
were taken on CCD (2068x4632) starting at 2012 Nov.~4 22:45~UT and 2012 Nov.~5
01:50~UT respectively. Data have been reduced using MIDAS package {\sc Echelle}.
Fifty echelle orders were extracted in the region of 4080 -- 7181~\AA\ with the
wavelengths calibrated via ThAr hallow-cathode lamp. The spectra normalization
was performed with the blaze function obtained from flat field exposures. Two
spectra of \pz\ were combined together with the allowance for the Earth
rotation. The average signal-to-noise ratio of the resulting spectrum ranges
from 25 in the blue to 110 in the red. 

At first glance the spectrum of \pz\ is typical for normal early K-stars. Closer
inspection shows however that there is a signature of the emission in the blue
part of the \Ha\ profile. Moreover, usually strong absorptions in K-stars 
(hydrogen lines, Mg\i\ 5170~\AA, and Na\i\ 5890~\AA) turn out to be not so deep
compared to normal red stars of similar temperature. The spectrum also reveals
He\i\ 5876~\AA\ (D$_3$) absorption that suggests the presence of the pronounced
chromosphere with the transition layer where the helium should be excited.
Chromospheric effects are likely responsible also for the relatively high
central intensity of strong absorption lines.

We do not confirm the presence of molecular features of TiO at 4760~\AA\ and
4950~\AA\ as well as MgH at 4780~\AA\ reported earlier
\citep{1989A&A...217..187P} which were the result of misinterpretation. The
spectrum synthesis reveals the atomic lines blends. We convolved our
spectrum with the Gaussian corresponding to the resolution of R=1200 in order
to compare \pz\ spectrum with the spectrum of \citet{1989A&A...217..187P} and
found no difference in the range of 4080--4960~\AA. Molecular 
line of TiO at 7055~\AA\ known as indicator of cool stellar spots is absent.
The spectrum in this region is essentially similar to those of normal red
giants.

The radial velocity derived from our spectrum is 27.3$\pm$0.4~\kms, a slightly
lower than 28.9$\pm$0.3~\kms\ reported by \citet{1998IBVS.4580....1S}. The
straightforward explanation of the difference  might be a binary nature of \pz.
Yet we did not find any signature of the secondary component in the \pz\
spectrum. 

\subsection{Galactic kinematics and metallicity}

The Hipparcos parallax of \pz\ \citep{2007A&A...474..653V} implies the distance
of 640$^{+1150}_{-250}$~pc. This distance when combined with the {\it Hipparcos}
proper motion and the radial velocity implies the galactic velocity components
($U,V,W$)=(17$\pm$3, 227$\pm$5, -18$\pm$17)~\kms. The galactic latitude of \pz\
is $-0^{\circ}.131$. This fact combined with $U,V,W$ velocity components imply
that \pz\ has almost galactic orbit in the galactic plane close to the solar
galactic radius. We conclude therefore that the metallicity of \pz\ should not
differ significantly from solar. We therefore assume the solar metallicity of
\pz; this assumption will be confirmed below by the detailed spectral analysis.

\subsection{Parameters of stellar atmosphere}

\begin{figure}
\centering
\includegraphics[width=0.45\textwidth,clip]{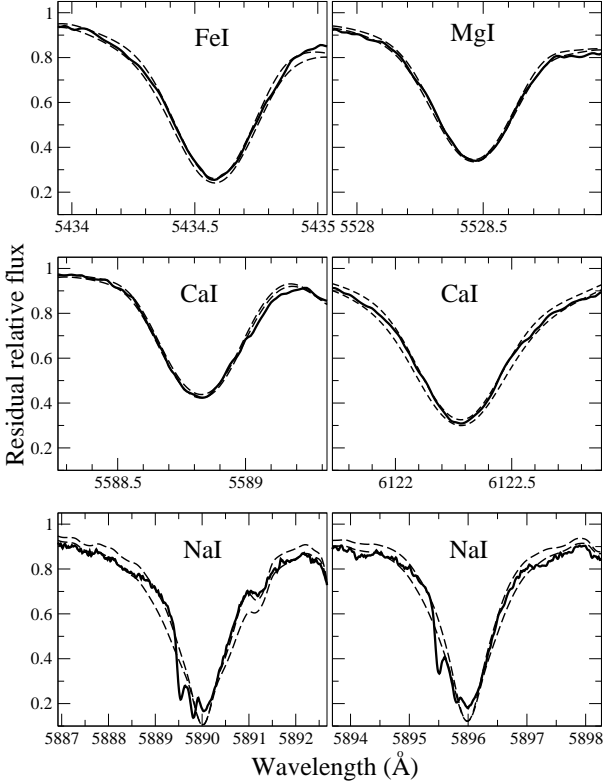}
\caption{Observed lines of several neutrals ({\it solid} line) compared to the
model spectra ({\it dashed}). For every profile we show two models which
correspond to the \lgg\ value that differ from the optimal by $\pm0.2$.}
\label{fig:wing}
\end{figure}

The rotation velocity derived from several unblended lines using {\sc BinMag3}
code\footnote{http://www.astro.uu.se/\~{}oleg/binmag.html}  
is \vs=10.5$\pm$0.6~\kms, very much close to
10.2$\pm$0.4~\kms\ found earlier \citep{1998IBVS.4580....1S}. This value exceeds
the typical rotational velocity of normal red giants but is consistent with the
fast rotation of active stars \citep{2007A&A...464.1101G, 2013CEAB...37..235O}. 
The macroturbulent velocity derived from profile analysis of 27 unblended line
of Fe\i, Ti\i, and Ni\i\ is \vm=5.2$\pm$0.5~\kms\ consistent with 
5.5$\pm$0.8~\kms\ reported by \citet{1998IBVS.4580....1S}, large for red 
dwarfs but
typical to red giants \citep{1982ApJ...262..682G, 1984ApJ...281..719G}. 

A standard prescription for the determination of the effective temperature makes
use of the requirement that the iron abundance derived from Fe\i\ lines should
not depend on the excitation potential ($E$). We selected 34 lines of Fe\i\ with
the equivalent width of 30-110~m\AA\ and excitation potential of
$0.91<E<4.99$~eV in the range of 5200--6700~\AA; this range permits us to avoid
strong molecular lines, telluric lines, and fringes. In the abundance
determination we exploit {\sc Width9} code written by V.~Tsymbal on the bases of
the R.~Kurucz code. Atomic and molecular data are extracted from VALD3 database
\citep{2011KIzKU.153...61R}. We derived the microturbulent velocity 1.4~\kms\
using the standard requirement that the Fe/H ratio should not depend on the
Fe\i\ line intensity. As a result we find the effective temperature
\Teff=4700$\pm$100~K and the iron abundance of [Fe/H]=0.0$\pm$0.1. The surface
gravity \lgg=2.9$\pm$0.2 is estimated from the profile fit of Fe\ii\ unblended
lines 5425.26, 6149.24, 6432.68, 6456.39~\AA\ using {\sc BinMag3} code. 

\begin{table}
\renewcommand{\tabcolsep}{1.0mm}
\caption{The parameters of strong spectral lines and the gravity estimation.}
\label{tab:logg}
\begin{tabular}{rrrrrrrr}
\hline
  \header{$\lambda$} &
  \header{Ion} &
  \header{$E$} &
  \header{log $gf$} &  
  \header{log $\Gamma_r$} &
  log $\frac{\Gamma_S}{N_e}$ &
  log $\frac{\Gamma_W}{N_H}$ &
  \header{\lgg} \vspace{1mm} \\
\header{\AA} & & \header{eV} & & s$^{-1}$ & s$^{-1}$ & s$^{-1}$ & \\
\hline
5434.5 & Fe\i & 1.011 & -2.122 & 7.140 & -6.220 & -7.749 & 2.93$\pm$0.11 \\
5528.4 & Mg\i & 4.346 & -0.498 & 8.720 & -4.460 & -6.979 & 2.47$\pm$0.18 \\
5588.8 & Ca\i & 2.526 &  0.358 & 7.853 & -6.072 & -7.538 & 2.68$\pm$0.19 \\
6122.2 & Ca\i & 1.886 & -0.316 & 7.860 & -5.320 & -7.189 & 2.75$\pm$0.19 \\
5889.9 & Na\i & 0.000 &  0.108 & 7.799 & -5.640 & -7.526 & 2.93$\pm$0.18 \\
5895.9 & Na\i & 0.000 & -0.194 & 7.798 & -5.640 & -7.526 & 2.87$\pm$0.15 \\
\hline
\end{tabular}
\end{table}

The independent diagnostics of gravity is provided by wings of strong spectral
lines. We selected four unblended lines of Fe\i, Mg\i, and Ca\i\  with the
residual flux of $<0.4$ and applied {\sc SynthV} code
\citep{2003IAUS..210P.E49T} adopting \Teff=4700~K to model the line profiles
adopting (Fig.~\ref{fig:wing}). The gravity \lgg\ is derived for every pixel and
the final value is obtained as the average for each line and presented in last
column of Table~\ref{tab:logg}. The table includes also atomic parameters
\citep{1999A&AS..138..119K}, viz., wavelength, ion, excitation potential of the
lower level, oscillator strength, radiation damping constant, Stark and van der
Waals broadening parameters calculated for 10\,000~K. 

In case of Na\i\ D$_1$ and D$_2$ we unable to fit the whole profile because
cores of these lines are affected by the chromosphere. We therefore concentrated
on the wings with the residual intensity $>0.3$ in the regions without the
interstellar lines. The \lgg\ values derived from Na\i\ lines
(Table~\ref{tab:logg}) agree with the \lgg\ derived from other neutrals and all
these values are consistent with \lgg\ inferred from Fe\ii\ lines. The overall
average value is \lgg=2.8$\pm$0.2. The effective temperature
\Teff=4700$\pm100$~K and gravity \lgg=2.8$\pm$0.2 suggests that \pz\ should be
classified as K2III giant. 

\subsection{Abundance of C, N, and Li}

Generally red giants show underabundance of carbon and overabundance of nitrogen
caused by the first dredge-up of products of CNO cycle
\citep{1981ApJ...248..228L}. The of CNO elements therefore can be an
independent signature of a red giant nature of \pz. The LTE {\sc BinMag3} code 
is used to determine the C and N abundance of \pz. We adopt parameters derived
above (\Teff=4700, \lgg=2.8, and $V_t$=1.4~\kms) and atomic/molecular data from
VALD3 database \citep{2011KIzKU.153...61R}. Analysis of molecular lines of C$_2$
at 5086~\AA\ and 5135~\AA, which are sensitive to the carbon abundance, results
in [C/H]=$-0.1\pm0.1$. Molecular lines of CH at 4835~\AA\ indicate somewhat
larger underabundance [C/H]=$-0.4\pm0.2$. The nitrogen abundance inferred from
CN molecular lines at 6250--6251~\AA\ is [N/H]=$0.15\pm0.20$. The slight
underabundance of carbon and overabundance of nitrogen with respect to the
solar composition are marginally consistent with the effect of the first
dredge-up and therefore are in line with the giant status of \pz.

\begin{figure}
\centering
\includegraphics[clip,width=0.47\textwidth]{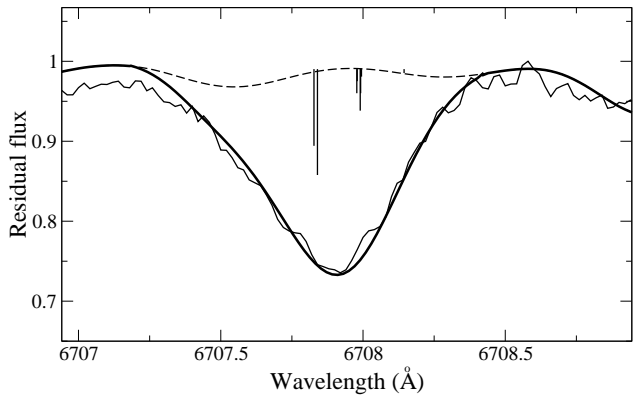}
\caption{Li\i\ 6707 \AA\ in the spectrum of \pz\ ({\it thin} line). The
synthetic spectrum is shown by {\it thick solid} line. The synthetic
spectrum without Li\i\ line is shown by {\it dashed} line. The
positions of lithium doublet components with hyperfine splitting are shown by
vertical lines.}
\label{fig:Li}
\end{figure}

The Li abundance in giants shows large scatter with
$A(\mbox{Li})=\log{(\mbox{Li/H})}-12\sim -1...~2.8$ in the range of \Teff$\sim$
4400-5000~K \citep{2011ApJ...741...26P} although the bulk of giants has
$A(\mbox{Li}) < 1.5$ and only 2\% have $A(\mbox{Li}) \geq 1.5$
\citep{1989ApJS...71..293B}. In \pz\ the Li 6707~\AA\ line is rather pronounced.
The contribution of blending lines of CN, Ce\ii, Sm\ii, Fe\i, and V\i\ is small
(Fig.~\ref{fig:Li}). To derive the lithium abundance we use {\sc BinMag3} code.
The hyperfine splitting and other atomic and molecular line contributions are
taken into account with the atomic data from VALD3. The resulting of LTE lithium
abundance is $A(\mbox{Li})=1.50\pm0.05$. Using a grid of non-LTE calculations
from \citep{2009A&A...503..541L} we recover the non-LTE correction
$A_{nLTE}-A_{LTE}=+0.19$~dex, so the corrected Li abundance is
$A(\mbox{Li})=1.69\pm0.05$. For consistency we compare our LTE result with
available LTE data. The value $A(\mbox{Li})=1.5$ turns out at the boundary
between giants  with relatively high Li abundance ($A(\mbox{Li}) \geq 1.5$) 
and normal K-giants ($A(\mbox{Li}) < 1.5$).

\subsection{Variation of radial velocity}

\begin{table}  
\caption{Radial velocity of \pz.\label{tab:Vr}}
\centering
\begin{tabular}{cccl}
\hline
\header{Set}&\header{JD}&\vr  & $\sigma$\vr  \\
            &               &\kms &~\kms    \\
\hline 
1 & 2447834.997 & 28.9  & 0.3  \\
2 & 2455860.607 & 29.58 & 0.22  \\
  & 2455861.620 & 28.94 & 0.28  \\ 
3 & 2456235.375 & 27.3  & 0.4   \\ 
4 & 2456598.584 & 24.18 & 0.23  \\ 
  & 2456602.597 & 28.35 & 0.30  \\ 
  & 2456603.545 & 29.09 & 0.27  \\ 
  & 2456604.639 & 29.28 & 0.34  \\ 
  & 2456605.709 & 29.36 & 0.29  \\ 
  & 2456606.609 & 30.72 & 0.24  \\ 
  & 2456607.613 & 31.98 & 0.28  \\ 
  & 2456611.558 & 29.14 & 0.28  \\ 
  & 2456612.592 & 29.05 & 0.37  \\ 
5 & 2456694.322 & 32.3  &  1.0  \\ 

\hline
\end{tabular}
\end{table}

\begin{figure}
\centering
\resizebox{0.95\hsize}{!}{\includegraphics[clip]{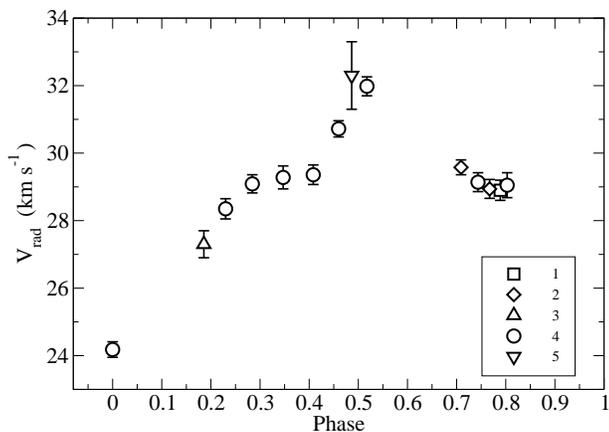}}
\caption{The radial velocity measurements convolved with period P=17.45~d.
Different sets (Table~\ref{tab:Vr}) are shown by different symbols.}
\label{fig:Phase}
\end{figure}

Only one measurement of the \pz\ radial velocity (RV) has been made formerly
\citep{1998IBVS.4580....1S} (\#1 in Table~\ref{tab:Vr}). Table columns give the
order number of the observational set, Julian day, radial velocity \vr\ and the
errors. The bulk of data (sets \#2 and \#4) we obtained using Radial Velocity
Meter (RVM) \citep{1987SvA....31...98T} installed at the Simeiz 1-m telescope  
of the Crimean Astrophysical Observatory. Zero point velocity was determined by
observations of several IAU velocity standards each night. The \#3 value was
described in Sec.~\ref{spectrum}. One point (\#5) was taken 2014 Feb. 5 using
MMSC echelle spectrograph (R=13000) at the 2-m telescope Zeiss-2000 of the
Terskol Observatory of Institute of Astronomy of Russian Academy of Sciences. 

Radial velocities of the major set \#4 show rise and fall of velocity between 24
\kms\ and 32~\kms\ with the maximum at about JD 2456607.6. Remarkably, the
velocity at JD 2456694.3, i.e., 86.7~d later (set \#5) $\approx32$~\kms\ is
similar to the maximum value. This leads to assumption that RV variations of
\pz\ are periodic, 86.7~d is an interval which consists of the integer number
of periods. We explored all possible periods in the range of $P\leq90$ days and
find only one optimum value P=17.45~d ($\approx(1/5)86.7$~d) which provides the
best fit for the folded data on the phase RV curve (Fig.~\ref{fig:Phase}). This
result indicates that the \pz\ radial velocity possibly varies periodically. 

\section{\pz\ distance}
\label{sec-param}

\subsection{Interstellar absorptions}

\begin{figure}
\centering
\includegraphics[clip,width=0.47\textwidth]{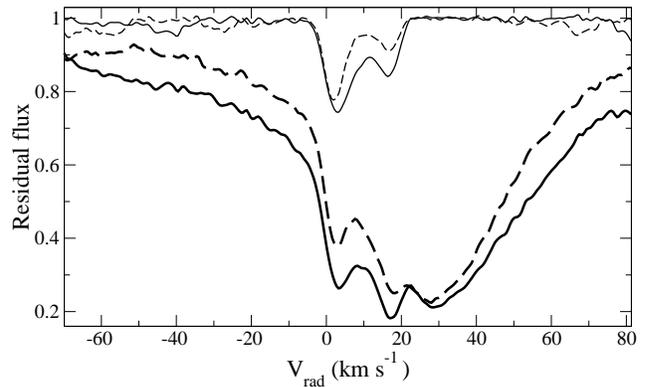}
\caption{Sodium D$_1$ ({\it thick dashed} line) and D$_{2}$ ({\it thick solid})
in the spectrum of \pz\ and extracted interstellar lines ({\it thin} lines).}
\label{fig:Na}
\end{figure}

The spectrum shows pronounced interstellar Na\i\ D$_1$ and D$_2$ absorptions
(Fig.~\ref{fig:Na}) which can be distance indicators. To extract the
interstellar Na\i\ lines we apply the mirror transform of red wing of the
stellar line with the control by synthetic spectrum. The extracted
interstellar lines are shown in Fig.~\ref{fig:Na}. The absorption shows two
components, A at 3.2~\kms\ and B at 17.5~\kms. The A component with the
equivalent widths of 36.2$\pm$2~m\AA\ and 28.5$\pm$2~m\AA\ in D$_2$ and D$_1$
lines respectively is obviously saturated, while the B component with the
equivalent widths of 20.4$\pm$2~m\AA\ and 11.0$\pm$2~m\AA\ is far from the
saturation. We use both of doublet components which makes the inferred column
density and broadening parameter more reliable
\citep[cf.][]{2012MNRAS.424.3145P}. The derived column density and broadening
parameter for A component are $N(\rm{NaI})=(9\pm2)\times10^{11}$~cm$^{-2}$ and
$b=0.5$~\kms\ respectively, while for B component
$N(\rm{NaI})=(1.1\pm0.1)\times10^{11}$~cm$^{-2}$ and $b=1.0$~\kms. 

The local interstellar medium is known to have of a low density and referred
to as local bubble with the radius of $\sim 100-150$~pc
\citep{2010A&A...510A..54W}. All sky survey of the Na\i\ interstellar
absorption \citep{1998A&A...333..101W} shows that towards the stars with 
distances $d<50$~pc the Na\i\ interstellar absorption is very weak with
$\log{N({\rm NaI})}<$11.0. The presence in the \pz\ spectrum of Na\i\
interstellar absorptions with $N(\rm{NaI})=(9\pm2)\times10^{11}$~cm$^{-2}$
therefore implies the distance to \pz\ of $>50$~pc. This disfavours the red
dwarf status for \pz\ in accord with the K2III classification based on the
spectral analysis.

\subsection{Stellar parameters and distance}
\label{subsec-param}

The derived stellar atmosphere parameters \Teff=4700$\pm100$~K and gravity
\lgg=2.8$\pm$0.2 can be combined with stellar evolutionary tracks to estimate
the mass. We employ evolutionary tracks \citep{2000A&AS..141..371G} for the
metallicity $Z=0.019$ to find $M$=1.5$\pm$0.5~$M_\odot$. This value alongside
with \lgg\ immediately provides us with the stellar radius of
$R=7.7\pm2~R_{\odot}$. The radius and \Teff\ imply the bolometric luminosity of 
$\log{L/L_\odot}=1.4\pm0.3$. For given \Teff, \lgg, and $B-V$ the Kurucz
models permits us to estimate the reddening $E(B-V)=0.08\pm0.02$~mag that agree
with value of 0.06~mag derived from IRFM. The angular diameter
($0.28\pm0.01$~mas) combined with the linear radius gives the estimation of
distance $d=250\pm70$~pc (or parallax $\pi=4.0\pm1.1$~mas). This value is by
factor 2.5 lower than distance estimated from {\it Hipparcos} parallax (640~pc).

\section{Chromosphere signatures}
\label{sec-chrom}

\begin{figure}
\centering
\includegraphics[width=0.45\textwidth,clip]{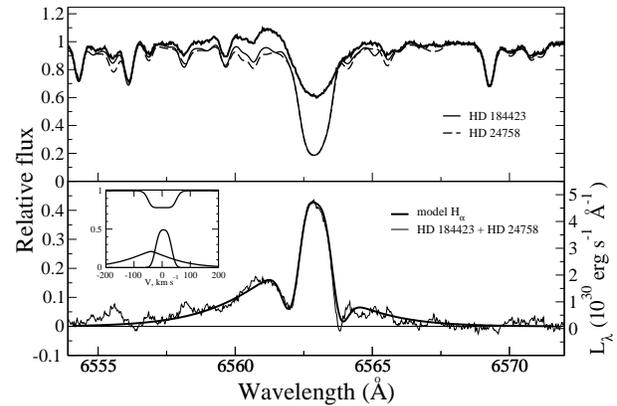}
\caption{\Ha\ profile in the spectrum of \pz\ ({\it thick solid} line) compared
to spectra of HD24758 and HD184423 ({\it top panel}). {\it Lower pane}: The
average residual ({\it thin} line) and model ({\it thick}). {\it Inset} shows 
model components.}
\label{fig:Ha}
\end{figure}

The spectrum of \pz\ shows the \Ha\ emission component presumably related to the
chromosphere. In Fig.~\ref{fig:Ha} the \pz\ spectrum in the \Ha\ band is 
shown alongside with the spectra of two normal red giants with similar
parameters: HD24758 (\Teff=4680~K, \lgg=2.75, [Fe/H]=0.11) and HD184423 
(\Teff=4680~K, \lgg=2.85, [Fe/H]=0.12) \citep{2011ARep...55..256P, 
2013AstL...39...54P}. Both comparison spectra were broadened with the rotation
velocity of \pz\ (10.5~\kms); telluric lines were removed from all the spectra.
Note that weak spectral lines in all three spectra are similar, while the \Ha\
of \pz\ is very much different: the absorption is shallow and in blue wing the
emission excess is apparent. The difference between the \pz\ spectrum and
average spectrum of the two other giants is shown in Fig.~\ref{fig:Ha} (bottom
panel). The monochromatic luminosity scale corresponds to $d=250$~pc and
$E(B-V)=0.06$~mag.

The residual \Ha\ emission consists of the central component with the luminosity
of $5.5\times10^{30}$~erg~s$^{-1}$ and broad component composed of the
blueshifted and redshifted emissions with the total luminosity of
$4.5\times10^{30}$~erg~s$^{-1}$. The overall \Ha\ luminosity (narrow + broad
emission) is $\approx10^{31}$~erg~s$^{-1}$, which corresponds to
$\approx\,10^{-4}$ of the \pz\ bolometric luminosity. We decompose the residual
\Ha\ profile into three components: narrow emission with FWHM $\approx
50$~\kms\ redshifted relative to photospheric spectrum by 5~\kms, broad
emission with FWHM $\approx 170$~\kms\ blueshifted by -40~\kms, and the central
saturated shallow absorption (Fig.~\ref{fig:Ha}, inset). The latter can be
described by the Gaussian absorption profile with the broadening of 32~\kms, the
central optical depth of $\tau_0=9$, and stellar surface covering factor of
0.22. This absorption is presumably related to enhanced hydrogen excitation in
the patchy chromospheric structure. The narrow \Ha\ emission arises in the
chromospheric plages likewise in RS~CVn stars \citep{2006A&A...446.1129B}.
The origin of the broad \Ha\ component which is usually seen in RS~CVn stars is
an unsettled issue. It probably originates from flaring activity
\citep{1997A&AS..125..263M}, although the corotating emission region or wind
are proposed as well \citep{1995AJ....109..350H}.

\subsection{He\i\ 5876 \AA\ absorption}

\begin{figure}
\centering
\includegraphics[clip,width=0.47\textwidth]{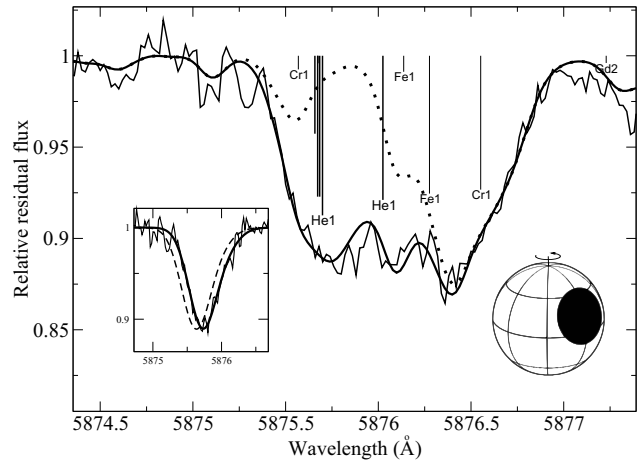}
\caption{He\i\ 5876 \AA\ in the spectrum of \pz\ ({\it thin} line). Synthetic
spectrum is shown by {\it thick solid} line. {\it Dotted} line shows synthetic
spectrum without He\i\ 5876 \AA. {\it Inset} shows the residual helium spectrum
({\it thin} line) and model spectrum without spot ({\it dashed} line) and with
the spot ({\it solid} line). The spot configuration is rendered in the
drawing ({\it right bottom}).}
\label{fig:He1}
\end{figure}

The He\i\ 5876~\AA\ absorption is a major component of the absorption feature in
the range 5875.3--5876.2~\AA\ (Fig.~\ref{fig:He1}). The He\i\ 5876~\AA\
absorption is commonly seen in spectra of active stars 
both dwarfs \citep{1997A&A...326..741S} and giants
\citep{1997A&AS..125..263M}. In the cool active star the helium line could form
only in the upper chromosphere. We recover the contribution of He\i\ absorption
as a residual from the ratio of the observed spectrum to the synthetic spectrum
of \pz\ (Fig.~\ref{fig:He1}, inset). The He\i\ absorption is redshifted by
4.5~\kms\ relative to photospheric spectrum. Remarkably, this redshift is
similar to that of the narrow \Ha\ emission component (5~\kms) which suggests
that both lines form probably in the same large active region. 

The He\i\ 5876 \AA\ redshift can be explained in the model of a large spot of
the excited He\i\ on the rotating star with the spot approaching the limb. For
the circular spot on a star with the inclination of the rotational axis
$i=70^{\circ}$ and rotational velocity of 11.4~\kms\ the optimal
parameters choice suggests the spot radius of 33.$^{\circ}$5  with the spot
center at the longitude of 43$^{\circ}$ relative to the central meridian and the
latitude of 20$^{\circ}$ counted from the rotational equator toward the
"visible" pole (Fig.~\ref{fig:He1}, inset). The adopted microturbulent velocity
is 8~\kms\ and the macroturbulent is 10~\kms. The limb darkening is treated in
the first Chandrasekhar approximation ($I(\theta) \propto 1/\sqrt{3} +
\cos\theta$). The model requires that the contribution of the rest of the
stellar disk in the He\i\ 5876~\AA\ absorption should not exceed 2\%. The direct
test of the spot model for the He\i\ 5876~\AA\ would be the detection of the
variable line shift with rotational phase.  

\section{Discussion}

\begin{figure}
\centering
\resizebox{0.95\hsize}{!}{\includegraphics[clip]{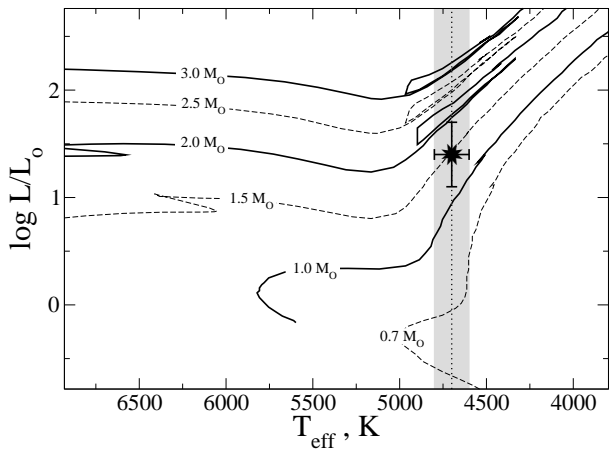}}
\caption{HR diagram with the \pz\ shown by star symbol. The range of temperature
of \pz\ is marked by {\it gray strip}.}
\label{fig:HR}
\end{figure}

Our primary goal was to establish the evolutionary status of \pz\ on the bases
of the detailed  photometric and spectral analysis. We inferred the effective
temperature and gravity which indicate that \pz\ is K2III giant. In combination
with the stellar evolutionary tracks the effective temperature and gravity
suggest the mass of \pz\ of $\approx1.5~M_{\odot}$. This in turn permitted us
to infer the radius, luminosity, and the distance of \pz; the latter estimate is
based on the stellar angular radius recovered by IRFM. The star resides in the
red giant branch of the HR diagram (Fig.~\ref{fig:HR}).

%% answer
Basically the presence of cool spots distorts the SED so the temperature
determined above by three methods should contain errors. Unfortunately, there is
no photometric data coeval with spectral ones. In order to estimate the
spottedness effects we therefore adopt the background stellar temperature of
4750~K and consider spots with the covering factor of 0.3 estimated by
\citet{2006A&AT...25..247A}. The spot temperature is assumed to be in the range
of 3500--4500~K. The fluxes are taken from the grid of models by
\citet{2003IAUS..210P.A20C}. The maximum change of ($B-V$) color caused by the
spottedness (+0.03~mag) is attained for $T_{spot}$=4000~K which results in the
reduction of the temperature by 83~K, i.e., within errors. The infrared fluxes
of spotted star are reduced by the factor of 0.9 so the angular diameter and the
distance determined by IRFM turn out larger by 5\%, also within errors. In this
case the new temperature is 4640~K, i.e., within errors. The temperature derived
from Fe\i\ lines is least sensitive to the stellar spottedness decreasing only
by 64~K. To summarize, the effective temperature of \pz\ may be in the range of 
4700--4800~K with the error about 100~K. The reddening derived in
Section~\ref{subsec-param} ($0.08\pm0.02$) is reduced for the spotty model to
$E(B-V)=0.06\pm0.03$.

Our spectrum reveals signatures of the robust \pz\ chromosphere demonstrated by
the \Ha\ emission and  He\i\ 5876~\AA\ absorption. The \Ha\ emission apparently
consists of two components: (i) narrow (FWHM$\approx50$~\kms\ and (ii) the
broad emission with FWHM$\approx 170$~\kms\ blueshifted by -40~\kms. This kind
of \Ha\ profile with blueshifted broad component is rather common for active
RS~CVn-type giants, {\it viz.}, HK~Lac (K0III), IM~Peg (K2III)
\citep{2006A&A...446.1129B}, DM~UMa \citep{1995AJ....109..350H}. At present
there is no unequivocal model for the broad component. The wind is unlikely
because in this case the required emission measure can be attained only for the
very high mass loss rate $\dot{M}\sim7\times10^{-9}v_{200}~M_{\odot}$~yr$^{-1}$
(where $v_{200}$ is the wind velocity in units of 200~\kms. This is by a factor
$\sim50$ larger than the mass loss rate suggested by the modified Reimers law
\citep{2005ApJ...630L..73S}. The flaring activity accompanied by the high
velocity ejecta is the likely explanation. The broad \Ha\ line-emitting sites 
in this scenario can be thought of as an analogue of the solar eruptive
prominences. This scenario naturally accounts for the dominant blueshift of the
broad component. Indeed, in this case we are able to account for both
the variability of broad \Ha\ and the dominance of the blueshift 
due to the occultation effect.

We find it remarkable that the He\i\ 5876~\AA\ absorption redshift (4.5~\kms)
and the narrow \Ha\ emission (5~\kms) redshift are similar. We interpret this as
an outcome of the spot on the rotating star in which case the He\i\ absorption
and \Ha\ narrow emission form in similar region of the chromosphere. In fact,
DM~UMa (RS~CVn type) shows synchronized rotational modulation of narrow \Ha\ and
He\i\ D$_3$ \citep{1995AJ....109..350H}. In this respect a question arises,
whether the active region responsible for \Ha\ and He\i\ D$_3$ is located over
the photospheric spot responsible for the photometric variations. At the moment,
simultaneous spectral and photometric observations are absent and this question
cannot be answered. It is instructive to estimate the spot size and the spot
temperature contrast required in order to account for the photometric variation
of 0.07~mag. We consider two cases: dark spot and bright spot assuming the
stellar temperature of 4700~K and the same spot radius $55^{\circ}$. We find
that the dark spot temperature should be 3800~K, while the bright spot 5200~K.

\begin{figure}
\centering
\includegraphics[width=0.45\textwidth]{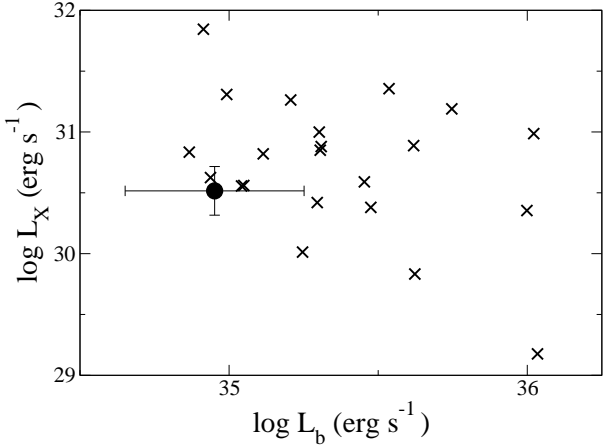}
\caption{\pz\ ({\it dot with error bars}) in the $L_x$ vs. $L_b$ scatter plot
for active giants. Data are from \citep{2007A&A...464.1101G}.}
\label{fig:xray}
\end{figure}

The archive {\em IUE} spectrum of \pz\ \citep{2000Ap&SS.273..155W} reveals
chromospheric lines of Mg\ii, O\i, Si\ii, He\ii, and C\iv\
(Table~\ref{tab:IUE}). The table contains also the observed integrated fluxes we
recovered from the line luminosities corrected for the extinction. Although the
chromospheric lines could be variable we note that the ratio of Mg\ii\
2800/\Ha=1.3 is comparable to the solar ratio Mg\ii\ 2800/\Ha$\approx2$
\citep{1979SSRv...24...71U}. We find that C\iv\ luminosity of \pz\
$\log{L(\mbox{C\iv})}\approx29.8$ (erg s$^{-1}$) is within the range 29.1--30.2
typical of RS~CVn binaries \citep{1993ApJ...413..333D}.

With the new distance \pz\ X-ray luminosity can be compared with other RS~CVn
stars. The {\it Einstein} count rate of \pz\ X-ray flux is $0.014\pm0.003$~cts  
\citep{1994HEAO2.C......0M}. To recover the X-ray unabsorbed flux we estimate
the interstellar hydrogen column density from the correlation between $N_{\rm
H}$ and $A_V$ \citep{2009MNRAS.400.2050G}, $N_{\rm H}=4\times10^{20}$~cm$^{-2}$,
and from the correlation between $N_{\rm H}$ and $N$(Na\i)
\citep{1985ApJ...298..838F}, $N_{\rm H}=4.9\times10^{20}$~cm$^{-2}$. Both values
are similar with the average of $N_{\rm H}=4.45\times10^{20}$~cm$^{-2}$.
Adopting this value and the typical coronal temperature of 1~keV for RS~CVn
stars \citep[e.g.][]{2007A&A...464.1101G} we find using HEASARC webPIMMS tools 
the unabsorbed flux of $(4.4\pm1.1)\times10^{-13}$~erg~cm$^{-2}$~s$^{-1}$ in
0.3--10~keV band. This corresponds to the X-ray luminosity
$L_X=(3.3\pm2.1)\times10^{30}(d/250{\rm pc})^2$~erg~s$^{-1}$. The found value
exceeds the X-ray luminosity of normal single red giants by a factor of
$\sim10^3$ \citep{1998A&A...335..591S}. On the other hand, the X-ray luminosity
of \pz\ gets into the range $\log{L_X}\approx 29-32$ (erg~s$^{-1}$) typical of
active  giants in binaries \citep{2007A&A...464.1101G} as demonstrated by the
scatter diagram (Fig.~\ref{fig:xray}). With regard to the X-ray luminosity \pz\
turns out typical RS~CVn type active giant. The attribution of \pz\ to the
RS~CVn class has been first proposed by \citet{1998IBVS.4580....1S}.

The classification of \pz\ as RS~CVn type star implies that the star could be a
binary \citep{1998IBVS.4580....1S}. We found variations of the radial velocity,
although the appearance of the radial velocity curve with the sharp maximum is
unusual for RS~CVn binaries. We therefore considered an alternative scenario:
the rotational modulation due to the star spot. Assuming that the photometric
period of 34.14~d is the rotational period the stellar radius of $7.7~R_{\odot}$
suggests the equatorial rotational velocity of $v=11.4\pm3.0$~\kms. With
\vs=10.5~\kms\ one finds then the inclination $\sin{i}=0.92\pm0.25$. These
parameters are used to find the optimal size of the spot assuming certain 
spot brightness contrast. It turns out that the model aimed to account for the
required line shift predicts the photometric variability with the amplitude of
$\approx0.45$~mag. This is at odds with the observed amplitude of only 0.07~mag.
We therefore reject this scenario. Note however that this conclusion does not
refer to the spot model for the redshift of chromospheric lines and spot model
for the photometric variability.

We are left thus with the conjecture that the velocity variations are caused 
by the orbital motion probably with some effects of spot superimposed on the 
smooth curve. In that scenario the suggested orbital period is $P=17.45$~d and 
the velocity amplitude of 3.9~\kms. Assuming zero eccentricity we come to the
mass function $f(M_2)=(M_1\sin{i})^3/(M_1+M_2)^2=1.08\times10^{-4}$. With the 
inclination $0.67<\sin{i}<1$ and the primary mass of $M_1=1.5~M_{\odot}$ the
mass function suggests the secondary mass of $0.064<M_2<0.097~M_{\odot}$. This
range is close to the brown dwarf limit ($\sim0.075~M_{\odot}$) with little hope
to detect the companion via the direct imaging (apparent magnitude
$\sim27$~mag). The only reasonable confirmation of the binary nature of \pz\ 
will be the periodic radial velocity curve based on a more complete data set. 

With the new classification \pz\ reveals the new quality as the RS~CVn variable 
with the longest photometric data set ($\sim100$~yr), the longest quasi-period
of $\sim50$~yr of activity and the largest amplitude of long-term variation
$\sim0.8$~mag among studied RS~CVn stars.  

\begin{table}
\label{tab:IUE}
\caption{Fluxes of emission lines in UV spectrum of \pz.}
\centering
\begin{tabular}{clccc}
\hline
  $\lambda$ &
  Ion &
  Flux &
  $L_{line}$ &  
  log $\frac{L_{line}}{L_{bol}}$ \\
\header{\AA} & & \header{erg~cm$^{-2}$~s$^{-1}$} 
& \header{erg~s$^{-1}$} & \\
\hline
1305 & O\i   & 6.1$\times$10$^{-14}$ & 8.1$\times$10$^{29}$ & -5.1 \\
1548 & C\iv  & 5.3$\times$10$^{-14}$ & 6.4$\times$10$^{29}$ & -5.2 \\
1640 & He\ii & 2.9$\times$10$^{-14}$ & 3.5$\times$10$^{29}$ & -5.4 \\
1815 & Si\ii & 2.8$\times$10$^{-14}$ & 3.4$\times$10$^{29}$ & -5.5 \\
2800 & Mg\ii & 1.2$\times$10$^{-12}$ & 1.3$\times$10$^{31}$ & -3.9 \\
\hline
\end{tabular}
\end{table}

\section{Conclusions}

The analysis of the photometric and spectral data leaves no doubt that the
active red star \pz\ is in fact an active  K2III giant with the chromosphere and
corona properties compatible with the typical RS~CVn star. This makes \pz\ the
object of a special interest because a long available photometric set of data
which shows the long-term variation of activity with the large amplitude. We
conjecture that \pz\ short-term variability with the period of 34.14~d is
related to the large active spot on the rotating star. We found similar
redshift of narrow \Ha\ emission and He\i\ D$_3$ absorption which are likely
related to the active spot on the rotating star. We found significant
variations of the radial velocity which could be caused by the orbital motion.

\section*{ACKNOWLEDGMENTS}

We are grateful to Ilya Sokolov for the spectral observation at Terskol
observatory. This work is supported by the grant "Nonstationary Phenomena in the
Universe" of the Russian Academy of Sciences. We thank the administration of the
Simeiz Section of the Crimean Astrophysical Observatory for allocating
observation time at the 1-m telescope. This work was partial supported by the
Russian Foundation for Basic Research (projects 11-02-00608 and 14-02-91153).

\bibliographystyle{mn2e}
\bibliography{paper}

\end{document}